%% file: prl.tex
\documentclass[aps,prl,twocolumn,showpacs,superscriptaddress,letterpaper]{revtex4}  % for review and submission
\usepackage{graphicx}
\usepackage{dcolumn}  
\usepackage{bm}       
\usepackage{amssymb}  

\begin{document}
%\pagewiselinenumbers

\newcommand{\ttbar}    {\ensuremath{t\bar{t}} }
\newcommand{\bbar}     {\ensuremath{b\bar{b}} }
\newcommand{\ppbar}    {\ensuremath{p\bar{p}} }
\newcommand{\qqbar}    {\ensuremath{q\bar{q}} }
\newcommand{\qqbb}     {\ensuremath{q\bar{q}b\bar{b}} }
\newcommand{\bbbar}    {\ensuremath{b\bar{b}} }
\newcommand{\ccbar}    {\ensuremath{c\bar{c}} }
\newcommand{\zjets}    {\ensuremath{Z+{\rm jets}} }
\newcommand{\invpb}    {\ensuremath{\rm pb^{-1}} }
\newcommand{\invfb}    {\ensuremath{\rm fb^{-1}} }
\newcommand{\invxx}    {\ensuremath{\rm fb^{-1}}}
\newcommand{\mbb}      {\ensuremath{m_{bb}}}
\newcommand{\mqq}      {\ensuremath{m_{qq}}}
\newcommand{\teststat} {\ensuremath{\Lambda} }
\newcommand{\et}       {\ensuremath{E_T} }
\newcommand{\pt}       {\ensuremath{p_T} }
\newcommand{\dr}       {\ensuremath{\Delta R} }
\newcommand{\me}       {\ensuremath{ |{\cal M}|^2 } }
\newcommand{\mh}       {\ensuremath{m_H } }
\newcommand{\cls}      {\ensuremath{CL_{s} } }
\newcommand{\h}        {\ensuremath{H } }
\newcommand{\vh}       {\ensuremath{VH } }
\newcommand{\fbinv}    {\ensuremath{\rm{fb}^{-1} } }
\newcommand{\gevcc}    {\ensuremath{\rm{{GeV}/}c^{2}  }}
\newcommand{\gev}    {\ensuremath{\rm{GeV} } }
\newcommand{\sigmasm}  {\ensuremath{\sigma_{\rm SM} } }
\newcommand{\sumet}    {\ensuremath{\sum E_T } }
\newcommand{\dzero}    {D\O}        % anyone know how they want this?

\bibliographystyle{apsrev}
%\leftline{Version 2.1 as of \today}

\title{ A Search for the Associated Production of the Standard-Model
  Higgs Boson in the All-Hadronic Channel }

\input{Jan_2008_Authors_Visitors1.tex}
%\author{The CDF Collaboration}
%\affiliation{URL: http://www-cdf.fnal.gov}

\begin{abstract}

We report on a search for the standard-model Higgs boson in $p\bar{p}$
collisions at $\sqrt{s}=1.96$ TeV using an integrated luminosity of 2.0 \invxx.
We look for production of the Higgs boson decaying
to a pair of bottom quarks in association with a vector boson $V$ ($W$ or $Z$) decaying to
quarks, resulting in a four-jet final state.
Two of the jets are required to have secondary vertices consistent
with B-hadron decays.
We set the first 95\% confidence level upper
limit on the $VH$ production cross section with $V(\rightarrow q\bar{q}/qq')H(\rightarrow b\bar{b})$ decay for Higgs boson masses
of 100-150 GeV/$c^2$ using data from Run II at the Fermilab Tevatron. For $m_H=120$ GeV/$c^2$, we exclude
cross sections larger than 38 times the standard-model prediction.
\end{abstract}


%for revtex
\maketitle

\date{\today}

% -- Introduction --

The standard model (SM) of elementary particle physics
includes a scalar Higgs ($H$) boson to explain the origins of
electroweak-symmetry breaking~\cite{higgs,englert}.
Direct searches for the Higgs scalar boson at the LEP collider~\cite{Barate:2003sz} have
constrained the Higgs boson mass ($\mh$) to be greater than
114.4 \gevcc\ at 95\% confidence level (CL).
For Higgs boson masses above this limit, the CDF and \dzero~experiments
at the Tevatron collider are currently performing
the most sensitive searches~\cite{Alcaraz:2006mx}.
Global fits to electroweak data~\cite{Collaboration:2008ub} indicate
a light SM Higgs boson, excluding $\mh > 154\ \gevcc~$ at 95\% CL. Searches
for a low-mass Higgs boson are thus particularly relevant.  For $\mh < 135\ \gevcc$, the
dominant decay mode is $H \rightarrow
\bbar$ ~\cite{Djouadi:1997yw}. While the dominant production modes are
direct $gg\rightarrow H$ and $q\bar{q} \rightarrow H$, the
$\bbar$ signature in this channel is overwhelmed by background from QCD $\bbar$ production.
Searches for events where the Higgs boson is produced
in association with a vector boson ($V$ = $W$ or $Z$) are
more promising. The $VH$ associated production cross section
is smaller by an order of magnitude than for direct
production, but identification of
the accompanying vector boson reduces the QCD
background, making searches for $VH$ the most sensitive
ones at low Higgs-boson mass.

So far, Tevatron Run II searches~\cite{higgsleptonic,zhll}
have used signatures where the $V$ decays to leptons. In this Letter we report on an analysis
of the channel in which the $V$ decays  to a $q\bar{q}$ pair resulting in
two jets.
Using data from 2.0 \invfb of integrated luminosity collected by the CDF experiment, we search
for four-jet events compatible with the $VH$ decay.
While
this channel has a large QCD background, it benefits from
the combined cross sections of $ZH$ and $WH$ production as well as the large
$V \rightarrow \qqbar/qq'$ branching ratio of about $70$\%~\cite{pdg}.  An analysis of this channel in Run I of the Tevatron~\cite{vhjjrun1} suggests strong potential.  This Letter presents the first analysis of this channel using data from Run II of the Tevatron; we find that uncertainties on the dominant background are larger than had been anticipated~\cite{vhjjextrap}.

% -- Detector and trigger --

% detector
% - copied verbatim from WH paper
% -  todo?: add some more information on calorimeter.

The CDF II detector~\cite{CDF1, CDF2} consists of a cylindrical magnetic spectrometer surrounded
by sampling calorimeters used to measure the energies of the jets.
Charged particle tracking is performed with silicon microstrip
detectors surrounded by a large cylindrical multilayer drift
chamber, immersed in a solenoidal magnetic field.
Planar drift chambers surround the calorimeters to detect muons.

% trigger

The data were collected using a three-level multi-jet online event
selection (trigger)~\cite{allhadtop}, originally designed for
hadronic top decays. To trigger a jet, in the first stage (level 1) a single
calorimeter tower was required with a transverse energy
($E_T$)~\cite{coordinates} of at least 10(20) GeV
for the data from the first (second) \invfb of integrated luminosity.  
At level 2, clusters of contiguous calorimeter towers were identified
and a fast online cluster energy measurement was performed.
Four clusters with $E_T > 15$ GeV were required. Additionally, the total transverse
energy, \sumet, was
required to exceed 125 (175) GeV for the first 0.4 (last 1.6) $\rm
fb^{-1}$ to reduce backgrounds from soft QCD jets. The thresholds were increased in the later periods to maintain an acceptable trigger
rate as the instantaneous luminosity increased over time. No
rejection is made by the level 3 trigger.

% Trigger efficiency studies

The trigger efficiency for the \vh signal is estimated using
a combination of {\sc pythia}~\cite{pythia} simulation and data.
Interaction of the final-state particles with the CDF\,II detector
is described by a {\sc geant}-based detector simulation~\cite{cdfsim}.
The data used to measure the efficiency were collected by triggers which required a single jet
with $E_T$ greater than 20 or 50 GeV.
Corrections to the simulated $VH$ trigger efficiency were
derived by comparing these data to the corresponding simulations.
The corrections account for imperfect simulation of the soft hadron
activity and jet finding in the trigger algorithm and differences in the energy scale at the trigger level between data and simulation. These corrections result in a
relative  reduction of the estimated efficiency for the $VH$ signal
by $\sim 20\%$.
A systematic uncertainty of 7\% (relative) on the trigger efficiency
is derived by comparing the corrections found in data with different
single-jet energy thresholds, and in different data periods.  The uncertainty of the trigger acceptance due to initial- and final-state radiation, the jet energy scale and the parton distribution functions
are handled separately.
The overall trigger efficiency for the $VH$ signal (with $\mh=120~\gevcc$)
is $33 \pm 2\%  (17 \pm 1\%)$ for the \sumet
threshold of $125(175)$ \gev.

In the final offline selection, jets are identified as energy
depositions in the calorimeters by the {\sc jetclu}~\cite{jetclu}
algorithm with a clustering radius of $0.4$ in azimuth-pseudorapidity space.
The reconstructed jet energies are corrected to give the best
estimate of the energy of each quark, including effects of calorimeter
response, multiple $p\bar{p}$ interactions, the underlying event, and
energy deposited outside the clustered jet~\cite{jes}.
Jets originating from $b$-quarks are identified, or `$b$-tagged', by the {\sc secvtx}
~\cite{secvtx} algorithm, which searches for a secondary vertex that results
from the displaced decay of a B-hadron.

\begin{figure}
\includegraphics[width=8cm]{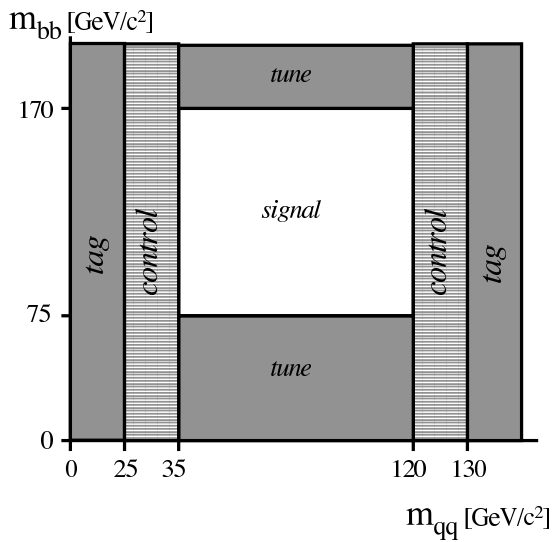}
\caption{Regions in the plane of $m_{bb}$, invariant mass of the two
$b$-jets, and $m_{qq}$, invariant mass of the two other jets. The {\it
tag} and {\it tune} regions are used to
define and tune the tag rate function used to predict
the background contribution from QCD $b\bar{b}$ production in the {\it signal} region.  The {\it
control} region is used to estimate a systematic uncertainty on the
interpolation of the tag rate function into the {\it signal} region.}
\label{fig:regions}
\end{figure}

Events compatible with the $VH\rightarrow qqbb$ signature
are selected by requiring at least       
four jets with $|\eta|<2.4$ and $E_T > 15$ GeV in which exactly two of the             %  AA     
jets are $b$-tagged.
The invariant mass of the $b$-tagged jets, $m_{bb}$, is required to exceed $75$\,GeV/$c^{2}$.
The invariant mass of the remaining leading two $q$ jets, $m_{qq}$, is required to be compatible
with the $W$ or $Z$ mass: $ 35 < \mqq < 120 $ \gevcc.
The di-jet mass resolution in the relevant invariant mass range is
of the order of 15 GeV/$c^{2}$, so the $WH$ and $ZH$ channels cannot be distinguished.
We refer to this as the {\it signal} region; see Fig.~\ref{fig:regions}.
Events in other
regions of the ($\mbb$, $\mqq$) plane and events with at least one
$b$-tag are used to model the QCD background to the
$VH\rightarrow qqbb$ signature.  
Events with identified isolated leptons are removed from the sample to
eliminate overlap with leptonic $W,Z$ decay channels and reduce the background
from $t\bar{t}$.
The combined trigger and selection efficiency for $VH \rightarrow qqbb$ events for the entire data-taking period varies from $\sim 1\%$ to $\sim 4\%$
for Higgs boson masses between 100 and 150 GeV/c$^{2}$.                 % wrt generator level!               

% Matrix Element Q

The dominant background to the $qqbb$ final state is
QCD multijet production. In order to distinguish between signal and background events,
we use the log-likelihood ratio
\[ Q = \log\left(  \frac{ P(\mathbf{x}| WH) + P(\mathbf{x}| ZH)}{P(\mathbf{x}| QCD)} \right), \]
where $\mathbf{x}$ is the vector of measured jet momenta of the
four highest $E_T$ jets,
and $P(\mathbf{x} | WH )$, $P(\mathbf{x} | ZH )$ and
$P(\mathbf{x} | QCD )$ are the likelihoods of observing
the event $\mathbf{x}$ for the $WH,ZH$ and QCD processes
respectively.  
The likelihoods are calculated by convoluting the differential       % convoluting <--> convolving ?
cross section as a function of the incoming and outgoing quark
momenta
for the processes with parameterized detector resolution functions,
and numerically integrating over the unmeasured magnitudes of the
quark momenta~\cite{memethod}.  $P(\mathbf{x} | \mathrm{process})$ is defined as                                                                         
\[ P(\mathbf{x} | \mathrm{process}) = \int d\Phi |\mathcal{M}_{\mathrm{process}}|^2
P_{tot} \]
\[ \times \prod_{j=1...4} T(E^j_{jet} | E^j_{quark})f_{p}f_{\bar{p}}, \]
where $d\Phi$ is the phase space of the unmeasured incoming and
outgoing quark momenta, $\mathcal{M}$ is the
matrix element, $P_{tot}$ is the probability density of the transverse
momentum of the process described by the matrix element,
$T(E_{jet} | E_{quark})$ is a transfer function which
parameterizes the probability to measure a quark of energy $E_{quark}$
as a jet with energy $E_{jet}$, and $ f_{p}$ and $f_{\bar{p}}$ are the
parton distribution functions~\cite{cteq} for the proton and anti-proton.  \
The jet directions are assumed to be measured perfectly, so that
the integration is over the magnitudes of the four quark momenta.

The matrix elements for $WH$ and $ZH$ are numerically calculated by the
{\sc alpgen}~\cite{alpgen} simulation. The matrix element
$\mathcal{M}_{gg\rightarrow ggbb}$ is used to describe the dominant background
process and is calculated by the {\sc madgraph}~\cite{madgraph} simulation.
However, these matrix elements do not describe initial state
radiation, which could result in non-zero total transverse momentum of the $VH$
system. The probability density of the transverse momentum, $P_{tot}$, is
extracted from simulated {\sc pythia} events that include radiation. The likelihoods $P(x|ZH)$,
$P(x|WH)$, and $P(x|QCD)$ are computed for \mh
between 100 and 150 \gevcc~with 10 GeV$/c^2$ intervals; values at
intermediate mass points are interpolated.  The transfer functions were
extracted from simulated events separately for $b$-jets and light-quark
jets, accounting for the reduced
jet energy for $b$-jets due to semileptonic decays.

% signal MC, background MC, QCD background model.
%
% - Swapped Discriminant part and sig/bg modeling part?
% - collected all stuff about MC together

% MC signal and background

Models of the $Q$ likelihood ratio distribution
are constructed for both signal and background events.
Backgrounds from $t\bar{t}$, single top,
and diboson production are modeled by {\sc pythia}. {\sc alpgen} is used to simulate the leading-order multiparton final state for the $W$ with heavy-flavor jets
background, while the hadronization and parton showering
are modeled by {\sc pythia}. Systematic uncertainties in the signal
acceptance, which includes trigger and selection efficiency, and the shape of the signal in $Q$ come from rates of initial- and final-state radiation, the jet energy
scale, the parton distribution functions, the trigger acceptance
and the $b$-tagging efficiency.  The 7\% trigger efficiency uncertainty is applied as a rate error
for all simulated samples. Uncertainties in the
cross sections of the background processes contribute to the systematic uncertainty in the background rate.

% QCD Background model    

A model for the primary background                               %% AA
is  constructed from the data using the background-dominated      %% Fucking Arthur Digby Sellers wrote
sample with at least one $b$-tagged jet. Each of the additional jets,    %% 156 episodes, Dude.
called a {\it probe} jet, is weighted by the probability  
for it to receive a $b$-tag, called the tag rate function (TRF).
For an event to contribute to the background model in the
signal region, the invariant mass of the tagged jet and the probe
jet must exceed 75 GeV/$c^2$ and the mass of the other two leading
jets, \mqq, in the event must be between 35 and 120 GeV/$c^2$.
Combinations for which \mqq\ is outside this window represent
an orthogonal set of probe-jets, which was used to measure
the TRF. In particular, the TRF is measured on
combinations where the mass of the other two jets are
incompatible with the vector boson masses, $m_{qq} < 25$ GeV/$c^2$ or $m_{qq} >
130$ GeV/$c^2$ (the region labeled \emph{tag} in Fig.~\ref{fig:regions}).

The tag rate is measured as a function of four
variables: the $\pt$ of the probe jet, the number of
tracks in the probe jet that traverse the silicon detector,
$\Delta R$
between the probe jet and the tagged jet, and the
invariant mass of the probe and the tagged jet. The TRF is
implemented using a four-dimensional histogram with bin-sizes
$14 ~{\rm GeV}/c~\times 2 \times 1 \times 15 ~{\rm GeV}/c^2$ respectively.
The TRF cannot be constructed
explicitly as a function of \mqq, as we use this variable
to interpolate from the sidebands into the signal region.
A small \mqq-dependence of the tagging rate may result from
correlations between \mqq\ and properties of the $b$-jets that
are not modeled by the TRF. The TRF is therefore further scaled
by the ratio of the observed \mqq\ distribution to the predicted
\mqq\ distribution in the low-mass region, $\mbb < 75$ \gevcc,
labeled {\it tune} region in Fig.~\ref{fig:regions}. The correction, a smooth function of \mqq, is of order 5\%.

We consider three sources of systematic uncertainty on the
shape of the $Q$ distribution for the QCD background.
The interpolation uncertainty accounts for possible differences in the TRF
between the regions where it was measured ({\it tag}) and applied ({\it signal}).
An alternative TRF is measured using events
with $ 25 < \mqq < 35 $ \gevcc\ or $ 120 < \mqq < 130 $ \gevcc
\ (labeled {\it control} in Fig.~\ref{fig:regions}). The difference
in the shapes of the predicted background distribution in $Q$ for the two TRFs
is treated as a systematic uncertainty.
The second source is due to uncertainty in applying the \mqq-tuning to the {\it signal} region.
An alternative tuning is derived using events with $\mbb > 170~$GeV$/c^2$, which is similarly background-dominated.
Finally, we estimate a mismodeling uncertainty
due to a possible limitation of the 4-dimensional
TRF parameterization to describe all the quantities
that affect the shape of the $Q$ distribution.  In a large
simulated \ttbar sample, we derive a TRF using events
in the {\it signal} region and use it to predict the number of double-tagged events                   % AA
in the same {\it signal} region. The difference in the {\it signal}           % AA
region between the $Q$ distribution for double-tagged events                  % AA
and TRF-weighted single-tag events is used to derive the mismodeling       % AA
uncertainty in the TRF method. This uncertainty describes any                 % AA
intrinsic failure of the TRF method to model $Q$ distributions,               % AA
independently of the details of the data sample.

\begin{figure}
\includegraphics[width=8cm,angle=90]{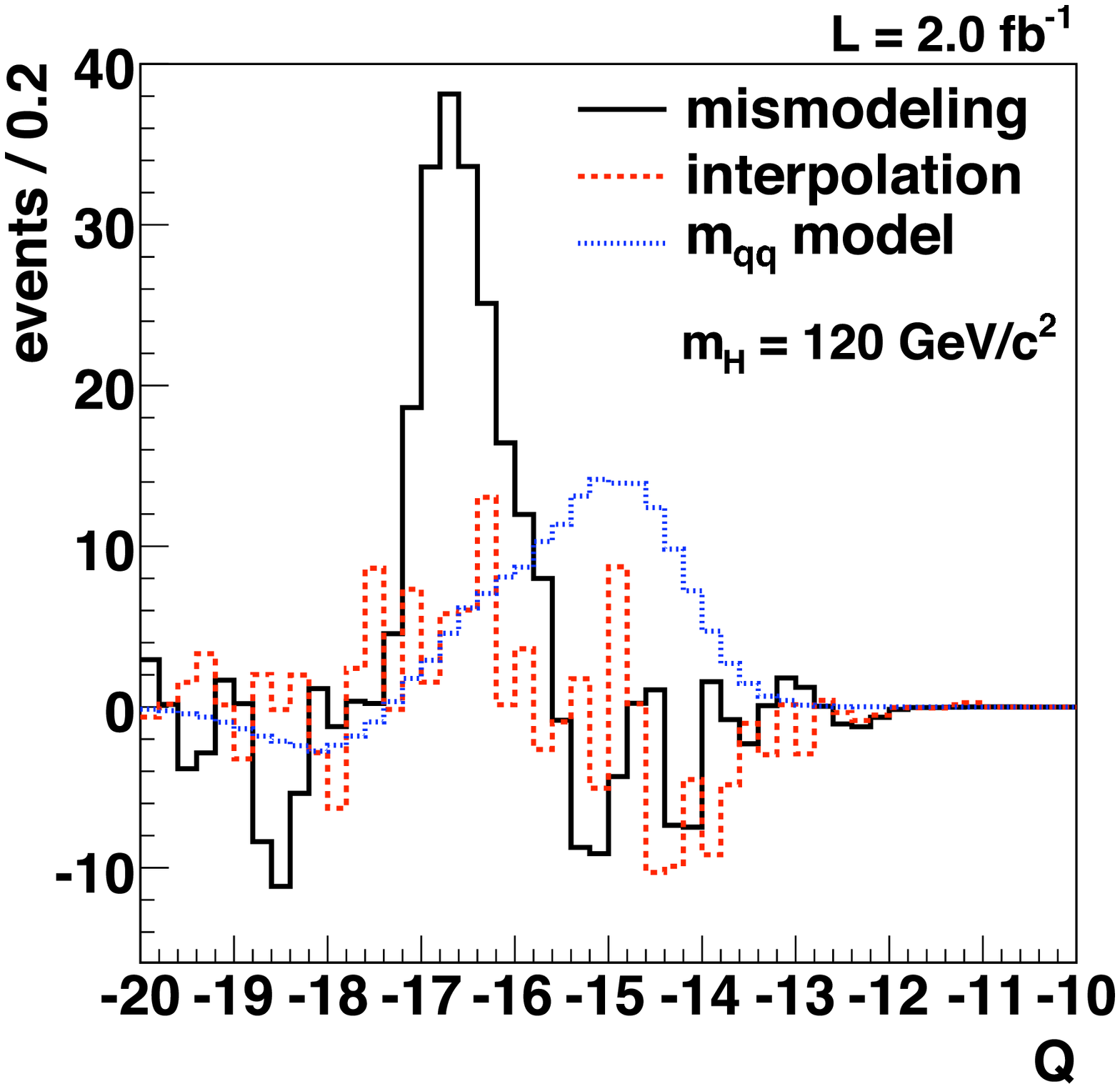}
\caption{ Systematic uncertainties on the relative number of
events expected for the QCD background model as a function of the
discriminant $Q$, from three sources described in the text.
%These distributions are scaled by a normally distributed random number and
%added to the background model used to generate the pseudo-experiments.
}

\label{fig:syst}
\end{figure}

The systematic shape uncertainties are shown in Fig.~\ref{fig:syst}, which
shows the expected deviation in the number of events within the uncertainties of the shape of the background model. In the region $Q \gtrsim -14$, where most of the        % AA
Higgs boson signal would be, the systematic uncertainty on the background          % AA
model is smaller than a few events per bin.                                  % AA

% Fit

\begin{figure}
\includegraphics[width=8cm]{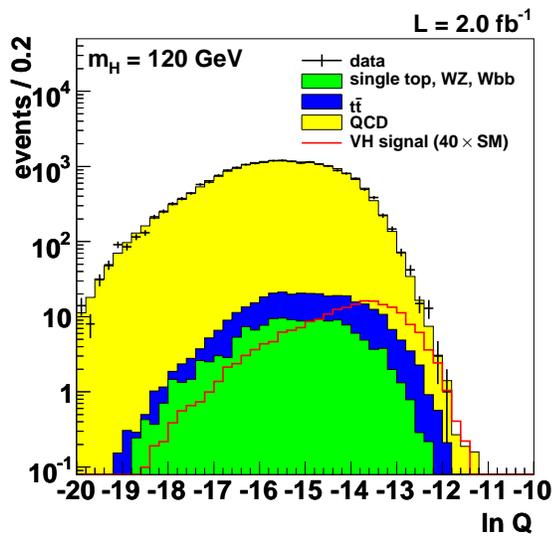}
\caption{ Predicted and observed distribution of the $Q$ discriminant
in the signal region in 2.0 fb$^{-1}$ of integrated luminosity. As $Q$ is a ratio of
likelihoods, its absolute scale is arbitrary.}
\label{fig:data}

\end{figure}

Figure~\ref{fig:data} shows the distribution of $Q$ for a signal
($m_H =120$ GeV$/c^2$),
the background contributions, and the observed
data. There is good agreement over the range of
$Q$ and no evidence of a $VH$ signal.

To test for the presence of a $VH$ signal in the data, a binned
likelihood of the distribution of the data in $Q$ is computed for the background-only
and the signal+background hypotheses. The ratio of these likelihoods
is the test statistic. The normalization of the QCD background model
is a free parameter that is fit to the data. The expected
distributions of the test statistic are derived from
pseudo-experiments that are generated from the signal
and background models, after varying them within the
systematic uncertainties. All systematic uncertainties are estimated as symmetric one standard-deviation
variations in the respective unknown nuisance parameter.  From these distributions,
limits on the $VH$ cross section are extracted using the
modified frequentist scheme~\cite{cls}.

% Results

Table~\ref{tab:lims} lists the expected and observed limits on
the $VH$ cross section expressed as a multiple of the SM
cross section, $\sigmasm$, for different Higgs boson masses.
The observed limits agree with the expected ones. For example, for a Higgs boson mass
of 120 GeV/c$^2$, the observed 95\% C.L. limit is
$38.2 \times \sigmasm$ while the expected limit is about $40 \times
\sigmasm$. At higher values of \mh the sensitivity
decreases due to the decreased Higgs boson production cross-section
and $\h \rightarrow \bbar$ branching fraction.
The systematic uncertainties on the background model          %AA
significantly affect the sensitivity: without these,          %AA
the expected limit would be 30\% lower.                        %AA

\begin{table}
\caption{ Number of predicted standard-model Higgs boson signal events and expected and observed 95\% confidence level upper limits on
the $VH$ production cross section, expressed as a multiple of the
standard model cross section for several Higgs boson masses. The final column
shows the observed limit on the cross section in $pb$.}
\ruledtabular
\begin{center}
\begin{tabular}{lrrrr}
$M_H$          & SM $VH$    & Expected lim.               & Obs. lim.                & Obs. lim. \\
(GeV/c$^{2}$)  &    $N$     &  $(\sigma/\sigma_{SM})$ &   $(\sigma/\sigma_{SM})$ & (pb)\\
\hline
100  & 7.2  &  $28.7_{-8.4}^{+12.7}$  &  29.7   & 13.9   \\
105  & 6.6  &  $33.7_{-9.9}^{+14.8}$  &  37.6   & 14.9   \\
110  & 6.4  &  $36.6_{-11.2}^{+15.8}$ &  38.7   & 13.2   \\
115  & 5.5  &  $36.8_{-10.4}^{+16.3}$ &  37.5   & 11.0   \\
120  & 5.2  &  $39.6_{-12.2}^{+16.8}$ &  38.2   & 9.66   \\
125  & 4.2  &  $46.8_{-14.3}^{+20.2}$ &  43.4   & 9.51   \\
130  & 3.5  &  $53.6_{-15.6}^{+22.9}$ &  48.0   & 9.13   \\
135  & 2.6  &  $80.2_{-23.2}^{+34.6}$ &  73.6   & 12.2   \\
140  & 2.0  &  $115_{-35.9}^{+51.7}$  &  107    & 15.5   \\
145  & 1.3  &  $176_{-52.7}^{+81.0}$  &  163    & 20.7   \\
150  & 0.86 &  $287_{-87.8}^{+127}$   &  261    & 28.9   \\
\end{tabular}
\label{tab:lims}
\end{center}
\end{table}

In summary, we report a limit on the production cross section of the standard-model Higgs
boson in association with a vector boson $V=W,Z$ with hadronic decays. Tighter limits are being obtained in the semileptonic decay channels. However, this is the first limit obtained in the difficult all-hadronic channel in Run II. We expect the analysis to be refined with time, and to be able to contribute to the overall Tevatron information on light Higgs production when all data of the
Tevatron Run II have been analyzed.

We thank the Fermilab staff and the technical staffs of the participating institutions for their vital contributions. This work was supported by the U.S. Department of Energy and National Science Foundation; the Italian Istituto Nazionale di Fisica Nucleare; the Ministry of Education, Culture, Sports, Science and Technology of Japan; the Natural Sciences and Engineering Research Council of
Canada; the National Science Council of the Republic of China; the Swiss National Science Foundation; the A.P. Sloan Foundation; the Bundesministerium f\"ur Bildung und Forschung, Germany; the Korean Science and Engineering Foundation and the Korean Research Foundation; the Science and Technology Facilities Council and the Royal Society, UK; the Institut National de Physique Nucleaire et
Physique des Particules/CNRS; the Russian Foundation for Basic Research; the Comisi\'on Interministerial de Ciencia y Tecnolog\'{\i}a, Spain; the European Community's Human Potential Programme; the Slovak R\&D Agency; and the Academy of Finland.

\bibliography{prl}

\end{document}

%% file: Jan_2008_Authors_Visitors1.tex
\affiliation{Institute of Physics, Academia Sinica, Taipei, Taiwan 11529, Republic of China} 
\affiliation{Argonne National Laboratory, Argonne, Illinois 60439} 
\affiliation{University of Athens, 157 71 Athens, Greece} 
\affiliation{Institut de Fisica d'Altes Energies, Universitat Autonoma de Barcelona, E-08193, Bellaterra (Barcelona), Spain} 
\affiliation{Baylor University, Waco, Texas  76798} 
\affiliation{Istituto Nazionale di Fisica Nucleare Bologna, $^t$University of Bologna, I-40127 Bologna, Italy} 
\affiliation{Brandeis University, Waltham, Massachusetts 02254} 
\affiliation{University of California, Davis, Davis, California  95616} 
\affiliation{University of California, Los Angeles, Los Angeles, California  90024} 
\affiliation{University of California, San Diego, La Jolla, California  92093} 
\affiliation{University of California, Santa Barbara, Santa Barbara, California 93106} 
\affiliation{Instituto de Fisica de Cantabria, CSIC-University of Cantabria, 39005 Santander, Spain} 
\affiliation{Carnegie Mellon University, Pittsburgh, PA  15213} 
\affiliation{Enrico Fermi Institute, University of Chicago, Chicago, Illinois 60637} 
\affiliation{Comenius University, 842 48 Bratislava, Slovakia; Institute of Experimental Physics, 040 01 Kosice, Slovakia} 
\affiliation{Joint Institute for Nuclear Research, RU-141980 Dubna, Russia} 
\affiliation{Duke University, Durham, North Carolina  27708} 
\affiliation{Fermi National Accelerator Laboratory, Batavia, Illinois 60510} 
\affiliation{University of Florida, Gainesville, Florida  32611} 
\affiliation{Laboratori Nazionali di Frascati, Istituto Nazionale di Fisica Nucleare, I-00044 Frascati, Italy} 
\affiliation{University of Geneva, CH-1211 Geneva 4, Switzerland} 
\affiliation{Glasgow University, Glasgow G12 8QQ, United Kingdom} 
\affiliation{Harvard University, Cambridge, Massachusetts 02138} 
\affiliation{Division of High Energy Physics, Department of Physics, University of Helsinki and Helsinki Institute of Physics, FIN-00014, Helsinki, Finland} 
\affiliation{University of Illinois, Urbana, Illinois 61801} 
\affiliation{The Johns Hopkins University, Baltimore, Maryland 21218} 
\affiliation{Institut f\"{u}r Experimentelle Kernphysik, Universit\"{a}t Karlsruhe, 76128 Karlsruhe, Germany} 
\affiliation{Center for High Energy Physics: Kyungpook National University, Daegu 702-701, Korea; Seoul National University, Seoul 151-742, Korea; Sungkyunkwan University, Suwon 440-746, Korea; Korea Institute of Science and Technology Information, Daejeon, 305-806, Korea; Chonnam National University, Gwangju, 500-757, Korea} 
\affiliation{Ernest Orlando Lawrence Berkeley National Laboratory, Berkeley, California 94720} 
\affiliation{University of Liverpool, Liverpool L69 7ZE, United Kingdom} 
\affiliation{University College London, London WC1E 6BT, United Kingdom} 
\affiliation{Centro de Investigaciones Energeticas Medioambientales y Tecnologicas, E-28040 Madrid, Spain} 
\affiliation{Massachusetts Institute of Technology, Cambridge, Massachusetts  02139} 
\affiliation{Institute of Particle Physics: McGill University, Montr\'{e}al, Canada H3A~2T8; and University of Toronto, Toronto, Canada M5S~1A7} 
\affiliation{University of Michigan, Ann Arbor, Michigan 48109} 
\affiliation{Michigan State University, East Lansing, Michigan  48824}
\affiliation{Institution for Theoretical and Experimental Physics, ITEP, Moscow 117259, Russia} 
\affiliation{University of New Mexico, Albuquerque, New Mexico 87131} 
\affiliation{Northwestern University, Evanston, Illinois  60208} 
\affiliation{The Ohio State University, Columbus, Ohio  43210} 
\affiliation{Okayama University, Okayama 700-8530, Japan} 
\affiliation{Osaka City University, Osaka 588, Japan} 
\affiliation{University of Oxford, Oxford OX1 3RH, United Kingdom} 
\affiliation{Istituto Nazionale di Fisica Nucleare, Sezione di Padova-Trento, $^u$University of Padova, I-35131 Padova, Italy} 
\affiliation{LPNHE, Universite Pierre et Marie Curie/IN2P3-CNRS, UMR7585, Paris, F-75252 France} 
\affiliation{University of Pennsylvania, Philadelphia, Pennsylvania 19104} 
\affiliation{Istituto Nazionale di Fisica Nucleare Pisa, $^q$University of Pisa, $^r$University of Siena and $^s$Scuola Normale Superiore, I-56127 Pisa, Italy} 
\affiliation{University of Pittsburgh, Pittsburgh, Pennsylvania 15260} 
\affiliation{Purdue University, West Lafayette, Indiana 47907} 
\affiliation{University of Rochester, Rochester, New York 14627} 
\affiliation{The Rockefeller University, New York, New York 10021} 

\affiliation{Istituto Nazionale di Fisica Nucleare, Sezione di Roma 1, $^v$Sapienza Universit\`{a} di Roma, I-00185 Roma, Italy} 

\affiliation{Rutgers University, Piscataway, New Jersey 08855} 
\affiliation{Texas A\&M University, College Station, Texas 77843} 
\affiliation{Istituto Nazionale di Fisica Nucleare Trieste/\ Udine, $^w$University of Trieste/\ Udine, Italy} 
\affiliation{University of Tsukuba, Tsukuba, Ibaraki 305, Japan} 
\affiliation{Tufts University, Medford, Massachusetts 02155} 
\affiliation{Waseda University, Tokyo 169, Japan} 
\affiliation{Wayne State University, Detroit, Michigan  48201} 
\affiliation{University of Wisconsin, Madison, Wisconsin 53706} 
\affiliation{Yale University, New Haven, Connecticut 06520} 
\author{T.~Aaltonen}
\affiliation{Division of High Energy Physics, Department of Physics, University of Helsinki and Helsinki Institute of Physics, FIN-00014, Helsinki, Finland}
\author{J.~Adelman}
\affiliation{Enrico Fermi Institute, University of Chicago, Chicago, Illinois 60637}
\author{T.~Akimoto}
\affiliation{University of Tsukuba, Tsukuba, Ibaraki 305, Japan}
\author{M.G.~Albrow}
\affiliation{Fermi National Accelerator Laboratory, Batavia, Illinois 60510}
\author{B.~\'{A}lvarez~Gonz\'{a}lez}
\affiliation{Instituto de Fisica de Cantabria, CSIC-University of Cantabria, 39005 Santander, Spain}
\author{S.~Amerio$^u$}
\affiliation{Istituto Nazionale di Fisica Nucleare, Sezione di Padova-Trento, $^u$University of Padova, I-35131 Padova, Italy} 

\author{D.~Amidei}
\affiliation{University of Michigan, Ann Arbor, Michigan 48109}
\author{A.~Anastassov}
\affiliation{Northwestern University, Evanston, Illinois  60208}
\author{A.~Annovi}
\affiliation{Laboratori Nazionali di Frascati, Istituto Nazionale di Fisica Nucleare, I-00044 Frascati, Italy}
\author{J.~Antos}
\affiliation{Comenius University, 842 48 Bratislava, Slovakia; Institute of Experimental Physics, 040 01 Kosice, Slovakia}
% ...

\collaboration{CDF Collaboration\footnote{With visitors from $^a$Universiteit Antwerpen, B-2610 Antwerp, Belgium, 
$^b$Chinese Academy of Sciences, Beijing 100864, China, 
$^c$University of Bristol, Bristol BS8 1TL, United Kingdom, 
$^d$University of California Irvine, Irvine, CA  92697, 
$^e$University of California Santa Cruz, Santa Cruz, CA  95064, 
$^f$Cornell University, Ithaca, NY  14853, 
$^g$University of Cyprus, Nicosia CY-1678, Cyprus, 
$^h$University College Dublin, Dublin 4, Ireland, 
$^i$University of Edinburgh, Edinburgh EH9 3JZ, United Kingdom, 
$^j$Universidad Iberoamericana, Mexico D.F., Mexico, 
$^k$University of Manchester, Manchester M13 9PL, England, 
$^l$Nagasaki Institute of Applied Science, Nagasaki, Japan, 
$^m$University de Oviedo, E-33007 Oviedo, Spain, 
$^n$Queen Mary, University of London, London, E1 4NS, England, 
$^o$Texas Tech University, Lubbock, TX  79409, 
$^p$IFIC(CSIC-Universitat de Valencia), 46071 Valencia, Spain,
$^x$Royal Society of Edinburgh/Scottish Executive Support Research Fellow,
$^y$Istituto Nazionale di Fisica Nucleare, Sezione di Cagliari, 09042 Monserrato (Cagliari), Italy 
}}
\noaffiliation

%% file: prl.bbl
\begin{thebibliography}{25}
\bibitem[{coo()}]{coordinates}
\bibinfo{note}{CDF uses a cylindrical coordinate system with the $z$ axis along
  the proton beam direction. Pseudorapidity is $\eta \equiv -\ln(\tan
  \theta/2)$, where $\theta$ is the polar angle, and $\phi$ is the azimuthal
  angle relative to the proton beam direction, while transerverse momentum $p_T
  = |p|\ \sin \theta$, and transverse energy $E_T = E\ \sin \theta$.}
\end{thebibliography}
